# Successful growth and room temperature ambient-pressure magnetic levitation of LK-99

Hao Wu[1,2]†, Li Yang[1,2]†, Bichen Xiao[1,2], Haixin Chang[1,2]*

[1]State Key Laboratory of Material Processing and Die & Mold Technology, School of Materials Science and Engineering, Huazhong University of Science and Technology (HUST), Wuhan 430074, China.

[2]Wuhan National High Magnetic Field Center and Institute for Quantum Science and Engineering, Huazhong University of Science and Technology (HUST), Wuhan 430074, China.

†These authors contributed equally to this work.

*Corresponding author. E-mail: hxchang@hust.edu.cn

**Abstract**

Recently, Sukbae Lee et al. reported inspiring experimental findings on the atmospheric superconductivity of a modified lead apatite crystal (LK-99) at room temperature (10.6111/JKCGCT.2023.33.2.061, arXiv: 2307.12008, arXiv: 2307.12037). They claimed that the synthesized LK-99 materials exhibit the Meissner levitation phenomenon of superconductors and have a superconducting transition temperature ($T_c$) higher than 400 K. Here, for the first time, we successfully verify and synthesize the LK-99 crystals which can be magnetically levitated with larger levitated angle than Sukbae Lee's sample at room temperature. It is expected to realize the true potential of room temperature, non-contact superconducting magnetic levitation in near future.

**Introduction**

With the advancement of science and technology, new superconducting materials are constantly being discovered. From conventional superconductors,[1] to iron-based superconductors (IBSCs),[2, 3] and then to copper oxide superconductors (cuprates),[4, 5] the discovery of new materials has broadened the categories of superconducting materials and continuously increases the superconducting transition temperature. Among them, the conventional superconductors mainly refer to high-purity niobium (Nb), high-purity vanadium (V), and niobium-titanium (Nb-Ti) alloys, etc.,[6-8] and their superconducting transition temperature is usually between several K to tens of K. Iron-based superconductors include LaFeAsO, $BaFe_2As_2$, etc.,[9-13] whose superconducting transition temperature is between 20 K and 55 K. Copper-based superconductors are superconducting materials based on copper oxides, and the superconducting transition temperature is above 90 K and up to 137 K,[14-17] which brings more possibilities for the application of superconducting materials. Currently, although some superconducting theories including Bardeen-Cooper-Schrieffer (BCS) theory and Bose-Einstein condensation (BEC) crossover theory have explained part of the superconducting mechanisms,[18, 19] the superconducting properties of some high-temperature superconductors still cannot be fully explained. The discovery of high temperature superconductors broadens the application prospect of superconducting materials, and sparks intense research activity to further improve the transition temperature of superconducting materials. Recently, Sukbae Lee et al. report possible room temperature ambient-pressure superconductor lead-apatite LK-99, and claim that the material has a $T_c$ higher than 400 K and has obvious Meissner levitation phenomenon.[20-22]

Here, we successfully for the first time verify and synthesize the LK-99 crystals which can be magnetically levitated with larger levitated angle than Sukbae Lee's sample at room temperature. It is expected to realize the true potential of room temperature, non-contact superconducting magnetic levitation in near future.

**Results and Discussions**

LK-99 samples with a composition of $Pb_{10-x}Cu_x(PO_4)_6O$ (0.9<x<1.1) were grown using the solid-state method similar to that reported by Sukbae Lee et al..[20-22] **Figure 1a** shows the synthesis temperature profiles of precursor Lanarkite ($Pb_2(SO_4)O$), precursor copper phosphide ($Cu_3P$) crystal and target product LK-99 from left to right, respectively. All the reactions are carried out under $10^{-2}$ Pa. The corresponding photograph and optical micrograph of an as-synthesized LK-99 crystal are shown in **Figure 1b** and **1c**. **Figure 1d** shows the crystal structure of LK-99, in which one of the four Pb(2) atoms is replaced by a Cu atom. **Figure 2a** and **Figure 2b** show the thermomagnetic curves (M-T, zero-field cooling (ZFC) and field cooling (FC) modes) in sample 1 (macroscopic gray-black bulk) and sample 2 (micron crystals screened by magnet repulsion, which is shaped in triangular with a side length of about 120 μm and a thickness of about 20 μm) measured in a physical property measurement system (PPMS, DynaCool, Quantum Design). For sample 1, the ZFC curves and FC curves show a diamagnetic transition at ~326 K and ~299K, respectively, which is similar to what Sukbae Lee reported earlier.[20] However, the diamagnetic transition temperature of micron crystal sample 2 screened by magnet repulsion is about 340 K, which is slightly higher than that of macroscopic sample, showing a higher purity, crystallinity and better Cu doping in the micron crystal sample 2. Such diamagnetic difference in different type samples indicates the fundamental potential superconducting mechanism with copper-oxygen induced electronic band structure changes in such phosphate oxides, as indicated by several theoretical studies.[23-25]

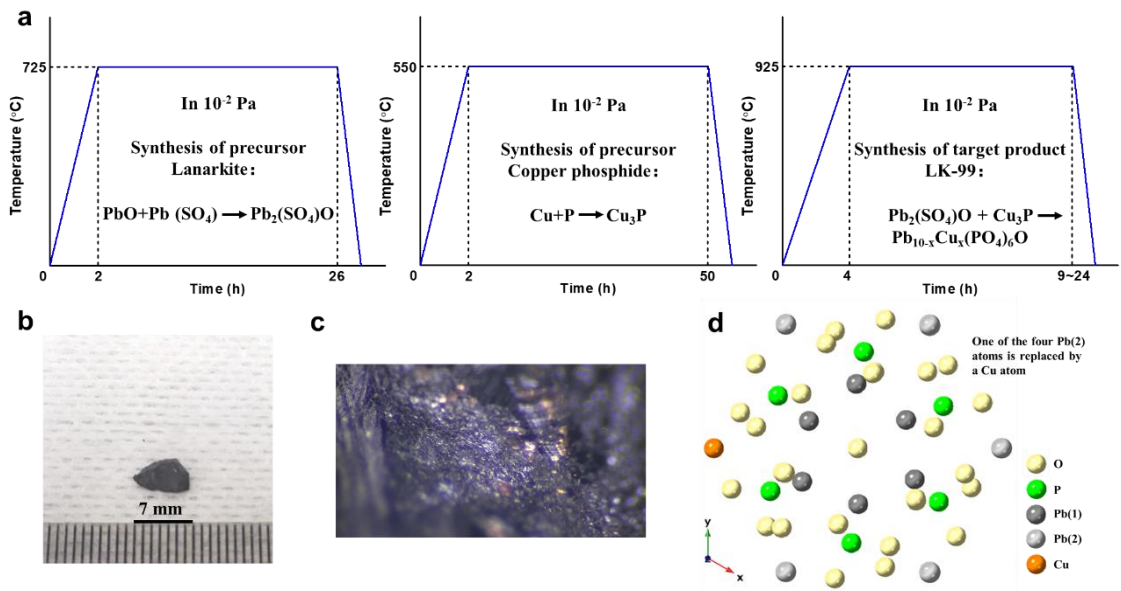

Figure 1. Synthesis of LK-99 samples. (a) Temperature profiles for the synthesis of the precursor Lanarkite and copper phosphide crystals and the target product LK-99. (b, c) Photograph (b) and optical microscopic image (c) of as-synthesized LK-99 sample. (d)

Schematic illustration of LK-99 crystal structure, in which one of the four Pb(2) atoms is replaced by a Cu atom.

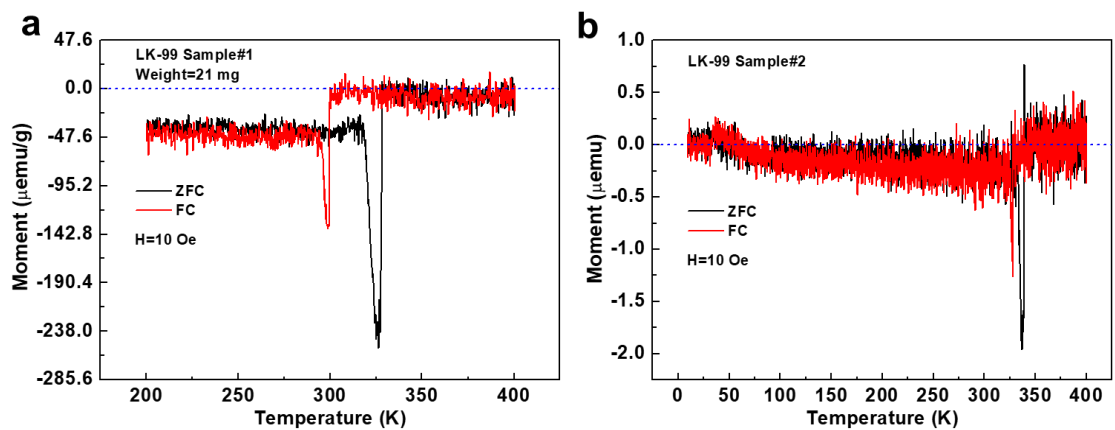

Figure 2. Diamagnetic transition of LK-99 samples. (a, b) Thermal evolution of the zero-field cooling (ZFC) and field cooling (FC) magnetization of the LK-99 samples 1 (a) and 2 (b) with 10 Oe magnetic field applied.

**Figure 3** further illustrates the levitation phenomenon of sample 2 measured at room temperature and atmospheric pressure, showing a rise and being completely perpendicular to the base when the ferromagnet is close to it, with a larger levitation angle than Sukbae Lee's sample at room temperature.[21] Video of the levitation is attached to Supplementary Material (see **Supplementary video 1**).

Moreover, we also performed an attraction test on sample 2 (**Figure 4**), which shows no response when attracted by a ferromagnet, excluding the effect of ferromagnetism. The corresponding video of the attraction is attached to Supplementary Material (see **Supplementary video 2**).

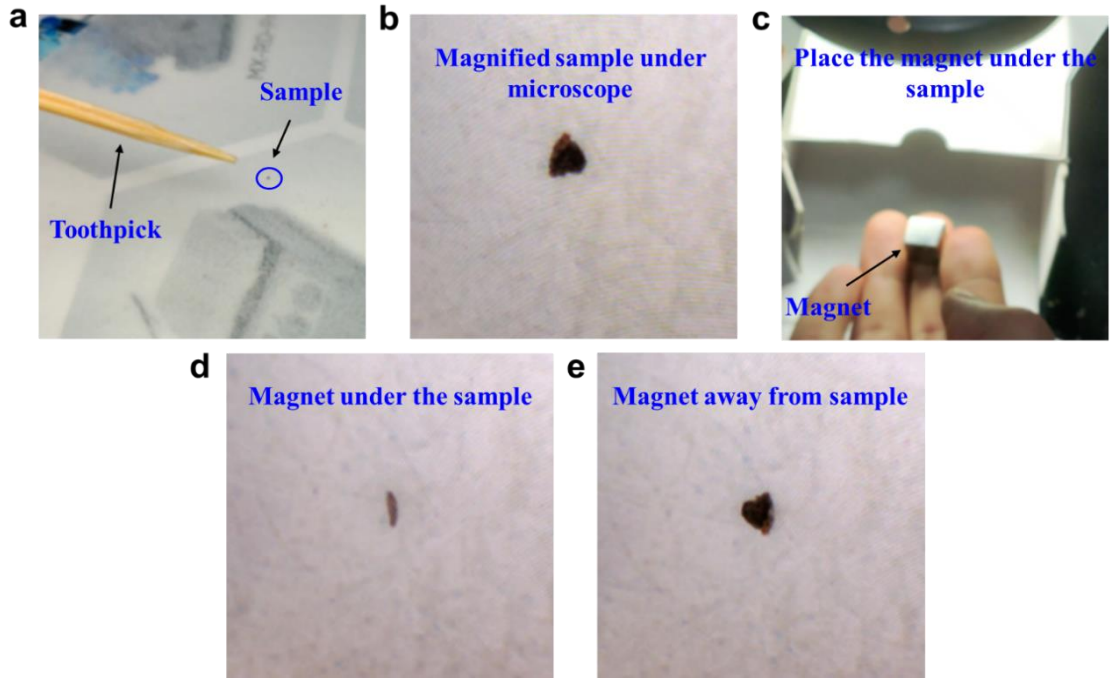

Figure 3. Diamagnetic levitation phenomenon for sample 2. Note that when the ferromagnet is near the sample, the sample rises and is completely perpendicular to the base. As the magnet moves away from the sample, the sample falls on the substrate.

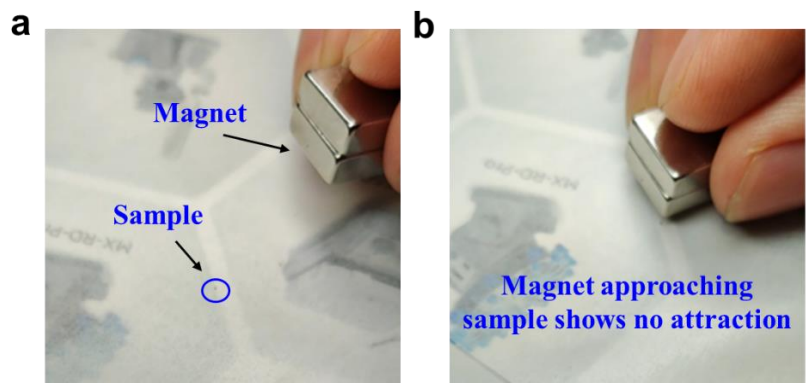

Figure 4. Exclude ferromagnetism of sample 2. Note that the sample did not respond when attracted to the magnet, ruling out ferromagnetism in sample 2.

**Conclusion**

We have successfully grown the LK-99 materials with consistent diamagnetism transition and large levitation angle at room temperature and ambient pressure. Our results show the importance of crystallinity and proper Cu doping, indicating the fundamental potential superconducting mechanism with copper-oxygen induced band changes in such phosphate oxides. We expect more consistent tests such as electrical tests in room temperature will show the great potential such phosphate oxides.

**Acknowledgments**

This work was supported by the National Key Research and Development Program of China (grant no. 2022YFE0134600), the National Natural Science Foundation of China (grant nos. 52272152, 61674063 and 62074061), the Natural Science Foundation of Hubei Province, China (grant no. 2022CFA031), the Foundation of Shenzhen Science and Technology Innovation Committee (grant nos. JCYJ20180504170444967 and JCYJ20210324142010030), and the fellowship of China Postdoctoral Science Foundation (grant no. 2022M711234).

**Competing interests**

The authors declare no competing interests.

**Data availability**

The data that support this study are available from the corresponding authors upon reasonable request.

# Supplementary Material to

# Successful growth and room temperature ambient-pressure magnetic levitation of LK-99

Hao Wu[1,2]†, Li Yang[1,2]†, Bichen Xiao[1,2], Haixin Chang[1,2]*

[1]State Key Laboratory of Material Processing and Die & Mold Technology, School of Materials Science and Engineering, Huazhong University of Science and Technology (HUST), Wuhan 430074, China.

[2]Wuhan National High Magnetic Field Center and Institute for Quantum Science and Engineering, Huazhong University of Science and Technology (HUST), Wuhan 430074, China.

†These authors contributed equally to this work.

*Corresponding author. E-mail: hxchang@hust.edu.cn

**Supplementary video 1：Meissner effect for sample 2.**

https://www.bilibili.com/video/BV14p4y1V7kS/?spm_id_from=333.999.0.0&vd_source=510b6eac8fec748481fb4b933932e80c

**Supplementary video 2：Exclude ferromagnetism of sample 2.**

https://www.bilibili.com/video/BV13k4y1G7i1/?buvid=XY81B1F843E91B69C1CC4563365B5FCE3A291&is_story_h5=false&mid=jLEqsyica5eHkvtMXQ2K1A%3D%3D&plat_id=193&share_from=ugc&share_medium=android&share_plat=android&share_source=QQ&share_tag=s_i×tamp=1690894807&unique_k=B6gawMH&up_id=7590247